\documentclass[5p,number]{elsarticle}

\usepackage{hyperref}
\usepackage[T1]{fontenc}
\usepackage[utf8]{inputenc}
\usepackage{textcomp} 
\usepackage{lmodern} 
\usepackage{amsmath}
\usepackage{amsfonts}
\usepackage{breakurl} 
\usepackage[switch]{lineno} 
\graphicspath{{./figures/}}
\biboptions{sort&compress}

\hypersetup{
    breaklinks = true,
    colorlinks = true,
    pdftitle = {}, 
    pdfauthor = {},
    pdfkeywords = {},
    linkcolor = blue,
    citecolor = blue,
    filecolor = black,
    urlcolor = magenta
}

\makeatletter 
\def\@author#1{\g@addto@macro\elsauthors{\normalsize%
    \def\baselinestretch{1}%
    \upshape\authorsep#1\unskip\textsuperscript{%
      \ifx\@fnmark\@empty\else\unskip\sep\@fnmark\let\sep=,\fi
      \ifx\@corref\@empty\else\unskip\sep\@corref\let\sep=,\fi
      }%
    \def\authorsep{\unskip,\space}%
    \global\let\@fnmark\@empty
    \global\let\@corref\@empty  
    \global\let\sep\@empty}%
    \@eadauthor={#1}
}
\makeatother

\newcommand{\yfeco}{Y(Fe$_{1-y}$Co$_y$)$_2$} 
\newcommand{\yfe}{YFe$_2$} 
\newcommand{\gdfe}{GdFe$_2$} 
 
\newcommand{\yco}{YCo$_2$}
\newcommand{\gdco}{GdCo$_2$}

\newcommand{\gdfeco}{Gd(Fe$_{1-y}$Co$_y$)$_2$}

\newcommand{\ygdfeco}{Y$_{1-x}$Gd$_x$(Fe$_{1-y}$Co$_y$)$_2$}
\newcommand{\ygdco}{Y$_{1-x}$Gd$_x$Co$_2$}
\newcommand{\ygdfe}{Y$_{1-x}$Gd$_x$Fe$_2$}

\begin{document}
\begin{sloppypar} 

\title{Monte Carlo calculations of Curie temperatures of\\  
\ygdfeco{} pseudobinary system}

\author[add1,add2,add3]{Bartosz Wasilewski}
\author[add1]{Miros\l{}aw Werwi\'nski\corref{cor1}}

\cortext[cor1]{Corresponding author} 
\ead{werwinski@ifmpan.poznan.pl}

\address[add1]{Institute of Molecular Physics, Polish Academy of Sciences,  M. Smoluchowskiego 17, 60-179 Pozna\'n, Poland}

\address[add2]{Institute of Mathematics, University of Szczecin, Wielkopolska 15, 70-451 Szczecin, Poland}

\address[add3]{Institute of Mathematics, University of Szczecin, Doctoral School, Mickiewicza 16, 70-383 Szczecin, Poland}

\begin{abstract}
%
%
The close-packed AB$_2$ structures called Laves phases constitute the largest group of intermetallic compounds.
In this paper we computationally investigated the pseudo-binary Laves phase system Y$_{1-x}$Gd$_x$(Fe$_{1-y}$Co$_y$)$_2$ spanning between the \yfe{}, \yco{}, \gdfe{}, and \gdco{} vertices.
While the vast majority of the Y$_{1-x}$Gd$_x$(Fe$_{1-y}$Co$_y$)$_2$ phase diagram is the ferrimagnetic phase, \yco{} along with a narrow range of concentrations around it is the paramagnetic phase.
%
We presented results obtained by Monte Carlo simulations of the Heisenberg model with parameters derived from first-principles calculations.
For calculations we used the Uppsala atomistic spin dynamics (UppASD) code together with the spin-polarized relativistic Korringa-Kohn-Rostoker (SPR-KKR) code.
%
%
From first principles we calculated the magnetic moments and exchange integrals for the considered pseudo-binary system, together with spin-polarized densities of states for boundary compositions.
Furthermore, we showed how the compensation point with the effective zero total moment depends on the concentration in the considered ferrimagnetic phases.
However, the main result of our study was the determination of the Curie temperature dependence for the system Y$_{1-x}$Gd$_x$(Fe$_{1-y}$Co$_y$)$_2$. 
Except for the paramagnetic region around \yco{}, the predicted temperatures were in good qualitative and quantitative agreement with experimental results,
which confirmed the ability of the method to predict magnetic transition temperatures for systems containing up to three different magnetic elements (Fe, Co, and Gd) simultaneously.
For the Y(Fe$_{1-y}$Co$_y$)$_2$ and Gd(Fe$_{1-y}$Co$_y$)$_2$ systems our calculations matched the experimentally-confirmed Slater-Pauling-like behavior of T$_C$  dependence on the Co concentration. 
For the Y$_{1-x}$Gd$_x$Fe$_2$ system we obtained, also in agreement with the experiment, a linear dependence of  T$_C$ on the Gd concentration.
In addition, on the example of Y$_{0.8}$Gd$_{0.2}$Co$_2$ ferrimagnet, we showed the possibility of predicting the non-trivial behavior of the temperature dependence of magnetization, confirmed by comparison with previous measurement results.
\end{abstract}

\date{\today}

\maketitle

\section{Introduction}
\begin{figure}[ht]
\centering
\includegraphics[clip,width=0.55\columnwidth]{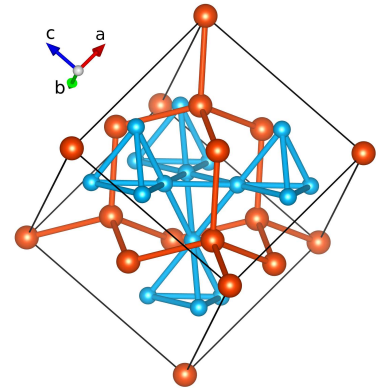}
\caption{\label{vesta} 
Crystal structure of \yco{} -- cubic MgCu$_2$-type Laves phase. 
Y atoms are shown in red, Co atoms in blue.
}
\end{figure}

\begin{figure}[t]
\centering
\includegraphics[clip,width=1.0\columnwidth]{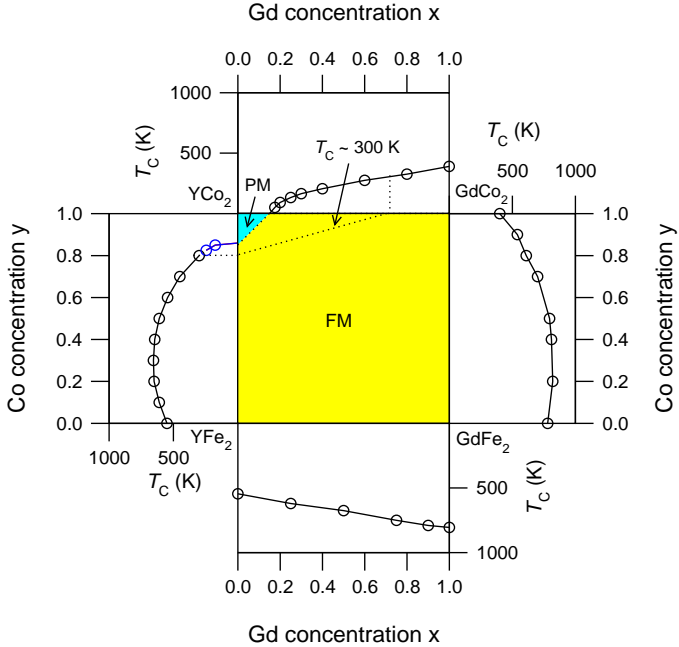}
\caption{ 
\ygdfeco{} phase diagram deduced from experimental Curie temperatures of boundary concentrations~\cite{guzdek_electrical_2012,buschow_magnetic_1970,burkov_magnetotransport_2003,burzo_paramagnetic_1978,kilcoyne_evolution_2000}.\label{fig:cross}
}
\end{figure}

The largest group of intermetallic compounds are the Laves phases~\cite{stein_structure_2004}.
They are binary close-packed structures with a formula AB$_2$, which 
are found in three different types - hexagonal MgZn$_2$-type (C14), cubic MgCu$_2$-type (C15), and hexagonal MgNi$_2$-type (C36), see Fig.~\ref{vesta}. 
In this work, we study theoretically the pseudo-binary cubic Laves phase \ygdfeco{}.
Its boundary cases are quite well known.
\gdfe{} and \gdco{} are ferrimagnets with Curie temperature (T$_C$) equal to 805~K~\cite{buschow_magnetic_1970} and 390~K~\cite{burkov_magnetotransport_2003}, respectively.
\yfe{} is a ferromagnet with Curie temperature equal to 545~K~\citep{buschow_magnetic_1970} and
\yco{} is an exchange-enhanced Pauli paramagnet~\cite{yamada_nmr_1980} that undergoes a metamagnetic transition in a field of 70~T at 10~K~\cite{goto_itinerant_1990,paul-boncour_metamagnetic_2019}.
In Fig.~\ref{fig:cross} we present a summary of the experimental Curie temperature -- concentration relationships for all boundary concentrations~\cite{guzdek_electrical_2012,buschow_magnetic_1970,burkov_magnetotransport_2003,burzo_paramagnetic_1978,kilcoyne_evolution_2000}.
We expect the \ygdfeco{} system to be paramagnetic (PM) in the vicinity of \yco{}, and a little further away to have $T_C$ close to room temperature, see the dotted line.
Alloying \yco{} with either Fe or Gd induces a paramagnetic-ferromagnetic transition at relatively low critical concentrations.

\begin{figure}[t]
\centering
\includegraphics[clip,width=1.0\columnwidth]{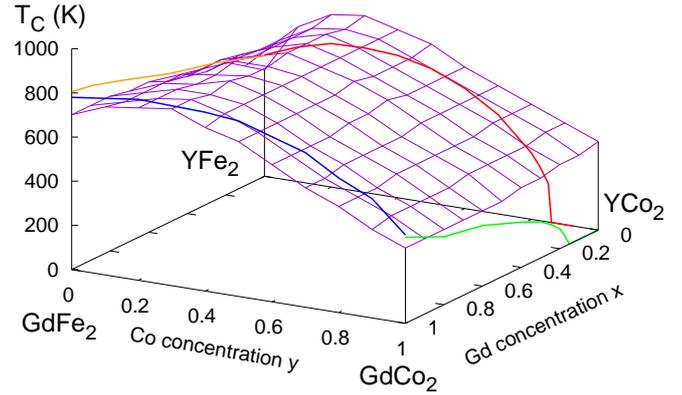}
\caption{
(Purple mesh) Curie temperature dependence of Gd and Co concentrations for \ygdfeco{}
determined from Monte Carlo simulations with parameters from the SPR-KKR code.
For comparison, the four curves (red, green, blue, and yellow) plotted using experimental results from Refs.~\cite{guzdek_electrical_2012,buschow_magnetic_1970,burkov_magnetotransport_2003,burzo_paramagnetic_1978,kilcoyne_evolution_2000}.
\label{fig:Tc}}
\end{figure}

\yfe{} has been identified as a hydrogen storage material~\cite{isnard_origin_2013,
paul-boncour_high_2017,
paul-boncour_metamagnetic_2019}.
\yco{} and its alloys have been considered for permanent magnet applications~\cite{kumar_permanent_2014}. 
For the system \ygdco{}, 
the magnetocaloric~\cite{pierunek_normal_2017}, 
thermopower~\cite{burkov_thermopower_2004}, and 
electronic transport~\cite{burkov_electronic_2011}
properties have been examined.
For the system \yfeco{}, 
electrical resistivity and M{\"o}ssbauer effect have been investigated~\cite{guzdek_electrical_2012}.
Moreover, rare-earth compounds with large localized magnetic moment, such as Gd, Tb or Er, are some of the most promising candidates for magnetic cooling among Laves phases~\cite{gschneidner_jr_recent_2005}.

%
Our previous experimental and theoretical studies on the \ygdfeco{} system have included such topics as:
effect of \yco{} doping~\cite{sniadecki_induced_2014},
magnetocaloric effects in \ygdco{}~\cite{pierunek_normal_2017},
Curie temperature of \yfeco{}~\cite{wasilewski_curie_2018},
structural disorder in \yco{}~\cite{sniadecki_influence_2018}, and
electronic specific heat coefficient of \yfeco{}~\cite{wasilewski_electronic_2019}.
%
%
In this work we show the Curie temperatures determined from first principles for pseudo-binary Laves phases \ygdfeco{}.
The two-site chemical disorder is modeled using Monte Carlo (MC) simulations. 
The $T_C$'s are obtained by fashioning the Heisenberg model Hamiltonian with MC simulations using parameters from first-principles calculations. 
We study the dependence of the exchange integrals on the interatomic distance and analyze the behavior of the total and partial magnetic moments calculated from first principles.
We also investigate the valence bands of the stoichiometric boundary compositions.
Furthermore, the MC method also allows us to predict the temperature dependence of magnetization ($M$) and magnetic susceptibility ($\chi$).
$T_C$ calculations were performed for systems containing up to four different elements simultaneously, including three magnetic elements.
Although we have previously presented $T_C$ dependence on concentration, e.g. for \yfeco{}~\cite{wasilewski_curie_2018}, in this work for the first time our calculations are based fully on MC simulations and not on the disordered local moment method as before.

\section{Calculations' details}

\begin{figure*}[t]
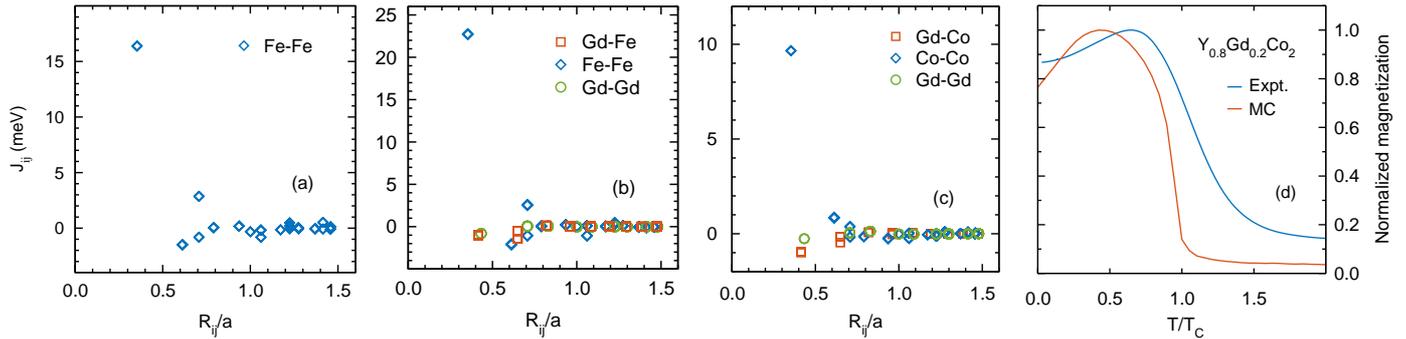

\centering
\includegraphics[clip,height=0.5\columnwidth]{x0y0_JXC.eps}\hspace{1mm}
\includegraphics[clip,height=0.5\columnwidth]{x1y0_JXC.eps}\hspace{1mm}
\includegraphics[clip,height=0.5\columnwidth]{x1y1_JXC.eps}\hspace{2mm}
\includegraphics[clip,height=0.5\columnwidth]{MagCurve.eps}
\caption{\label{fig:JXC} 
Exchange integrals as a function of normalized interatomic distance calculated using SPR-KKR for (a)~\yfe{}, (b)~\gdfe{}, and (c)~\gdco{}. 
For \yfe{} the moment on Y is fixed at zero, so only the Fe-Fe exchange integrals are shown.
(d) Normalized magnetization as a function of normalized temperature for Y$_{0.8}$Gd$_{0.2}$Co$_2$. 
Monte Carlo \textit{versus} experimental (zero-field cooled) results taken from Ref.~\cite{pierunek_normal_2017}.
}
\end{figure*}

One method for determining the Curie temperature is Monte Carlo simulations of the Heisenberg model. 
The simulations allow to determine the temperature dependence of magnetization and magnetic susceptibility.
The location of the peak in magnetic susceptibility allows us to determine $T_C$ with an accuracy of $\pm$10~K.
We obtained the values of the magnetic moments and the exchange integrals necessary to perform the MC simulations from first-principles calculations 
%
%
utilizing the spin polarized relativistic Korringa-Kohn-Rostoker (SPR-KKR) code, version 7.7~\cite{ebert_et_al._munich_nodate,ebert_calculating_2011}.
We used the generalized gradient approximation (GGA) in the Perdew, Burke, and Ernzerhof parametrization~\cite{perdew_generalized_1996} and the atomic-sphere approximation (ASA) in the fully-relativistic approach.
We used the coherent potential approximation (CPA)~\cite{soven_coherent-potential_1967} to simulate chemical disorder. 
We used the basis functions up to $l = 4$, a $45\times45\times45$ \textbf{k}-mesh, and 40 energy points. 
Exchange integrals were obtained using the method of Liechtenstein~\textit{et~al.}~\cite{liechtenstein_exchange_1984} with respect to the ferrimagnetic ground state.
Although in previous work we have used GGA~+~U  corrections to describe both  $d$~\cite{skoryna_xps_2016} and $f$~\cite{morkowski_x-ray_2011} valence electrons, due to the exploratory nature of this work we decided not to extend the current model beyond standard GGA. 
The effect of on-site Hubbard-type interactions on the values of the exchange integrals has recently been discussed elsewhere~\cite{keshavarz_magnetic_2020}.

%
For Monte Carlo simulations of the Heisenberg Hamiltonian we used the Uppsala atomistic spin dynamics (UppASD) code~\cite{skubic_method_2008,eriksson_atomistic_2017}.
The simulated system consisted of $\sim$13000 atoms with periodic boundary conditions. 
The radius of the exchange integrals cutoff sphere in the Heisenberg Hamiltonian was set to 1.5 lattice parameters. 
The exchange integrals were assumed to be temperature independent in the MC simulations.
The simulations were performed using classical (Boltzmann) statistics, but it is worth noting that it has recently been possible to use quantum (Bose-Einstein) statistics using the UppASD code~\cite{bergqvist_realistic_2018}. 
The magnetic moment on Y, since it is induced and expected to vanish with temperature, was set equal to zero in the MC simulations (first-principles calculations gave a moment of $\sim$0.5~$\mu_B$).
In the MC simulations, we used exchange integrals calculated with respect to the ferrimagnetic ground state. 
We justify this on the grounds that MC simulations of \gdco{} with exchange integrals obtained with respect to the paramagnetic ground state have yielded that material to be a paramagnet. 
This can be explained by the fact that the Co sublattice is metamagnetic in \gdco{}~\citep{gratz_physical_2001}.

%
Experimental lattice parameters were used to model the boundary compositions \yfe{} (7.36), \yco{} (7.22), \gdfe{} (7.38), and \gdco{} (7.24~$\AA$)~\cite{oesterreicher_studies_1967,guzdek_electrical_2012}. 
For intermediate \ygdfeco{} concentrations, we assumed a linear behavior of the lattice parameters.
The full range of Co and Gd concentrations were prepared with a step of 0.1, leading to a total of 121 cases considered ($11 \times 11$).
The space group and atomic positions of the C15 Laves phase are shown in Table~\ref{tab:wyckoff}.
The picture of the unit cell, generated using the VESTA code~\cite{momma_vesta_2008}, is shown in Fig.~\ref{vesta}.

\begin{table}[]
\centering
\caption{\label{tab:wyckoff} 
Atomic coordinates of the considered \yfe{}, \yco{}, \gdfe{}, and \gdco{} Laves phases, space group $Fd$-$3m$ (no. 227), origin choice two. 
}
\centering
\begin{tabular}{ccccc}
\hline \hline
atom  	& site	& $x$ 	& $y$ & $z$ \\
\hline
Y/Gd   	& 8$a$ 	& 1/8 	& 1/8   & 1/8 \\	
Fe/Co  	& 16$d$	& 1/2 	& 1/2   & 1/2 \\
\hline \hline
\end{tabular}
\end{table}

\section{Results and Discussion}

\begin{figure*}[t]
\centering
\includegraphics[trim = 0 0 0 0, clip,height=0.68\columnwidth]{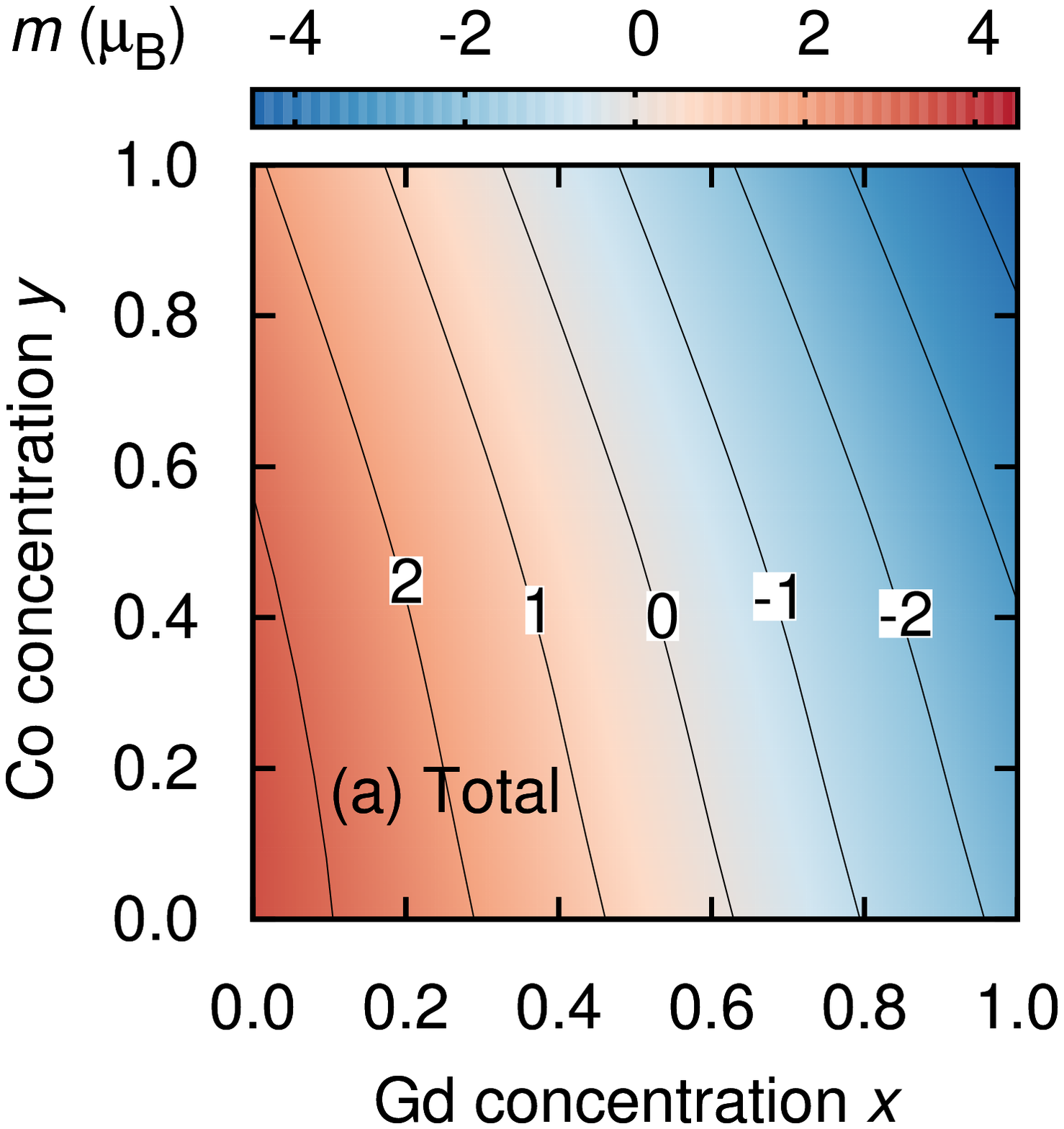}\hspace{1mm}
\includegraphics[trim = 0 0 90 0, clip,height=0.68\columnwidth]{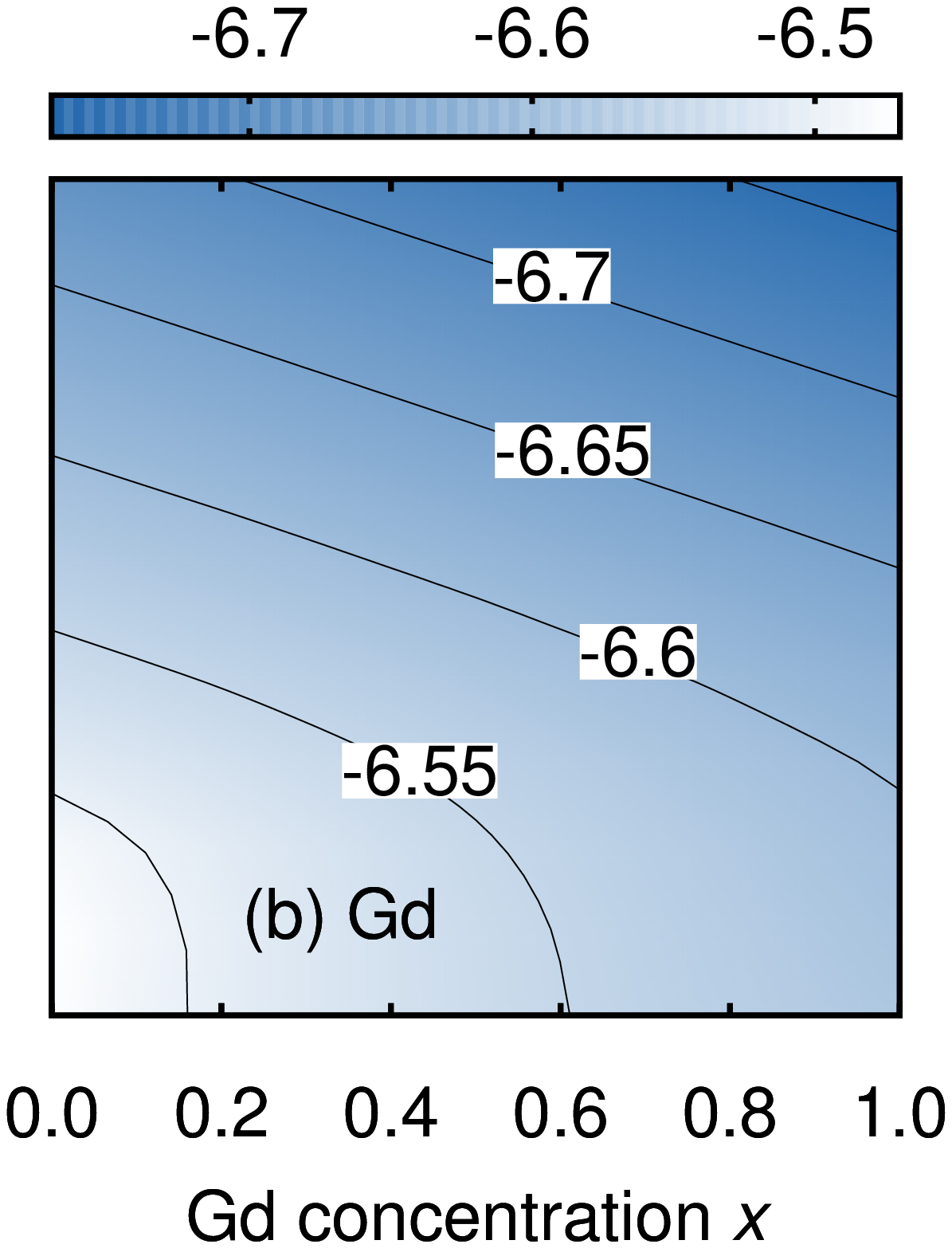}\hspace{1mm}
\includegraphics[trim = 90 0 0 0, clip,height=0.68\columnwidth]{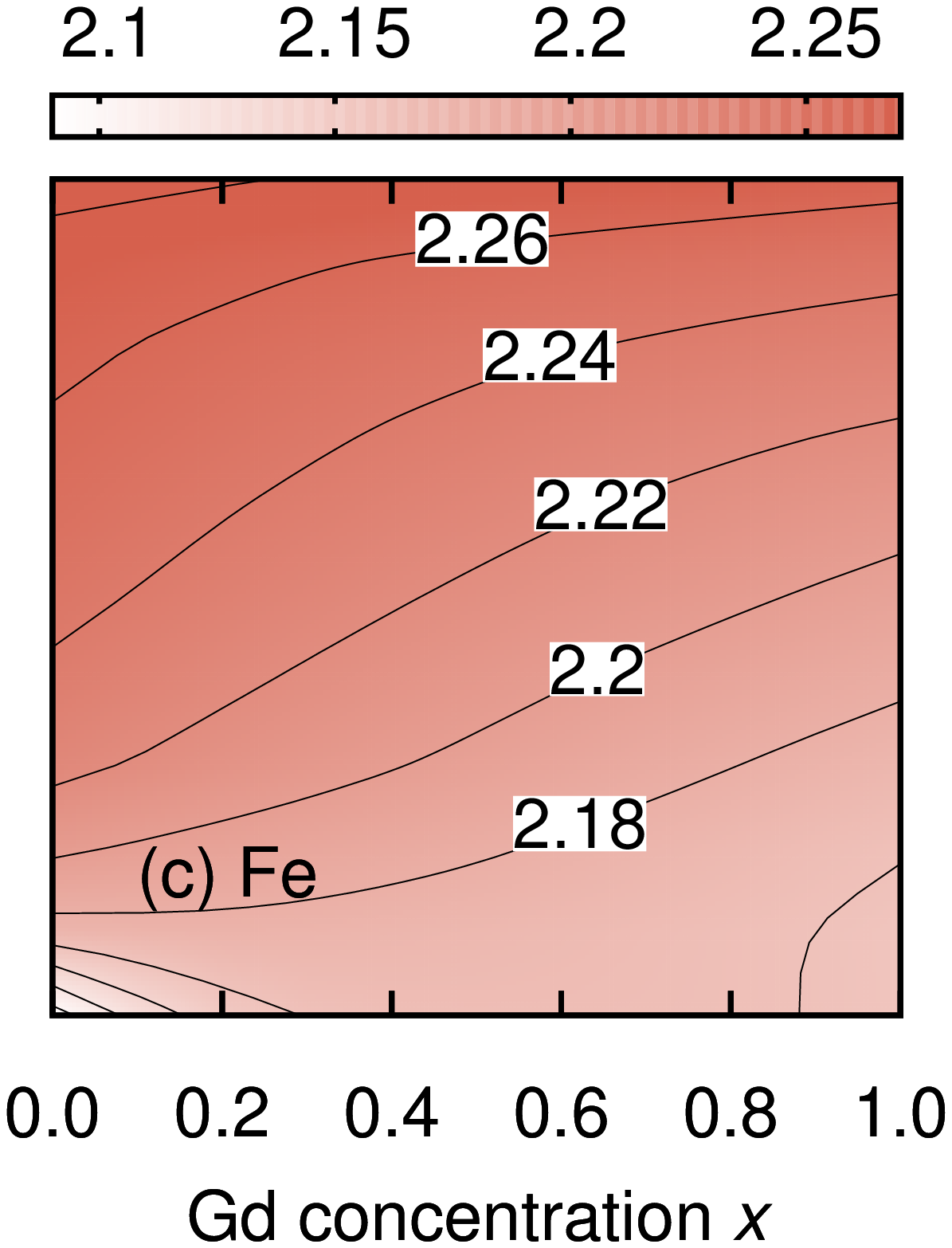}\hspace{1mm}
\includegraphics[trim = 0 0 90 0, clip,height=0.68\columnwidth]{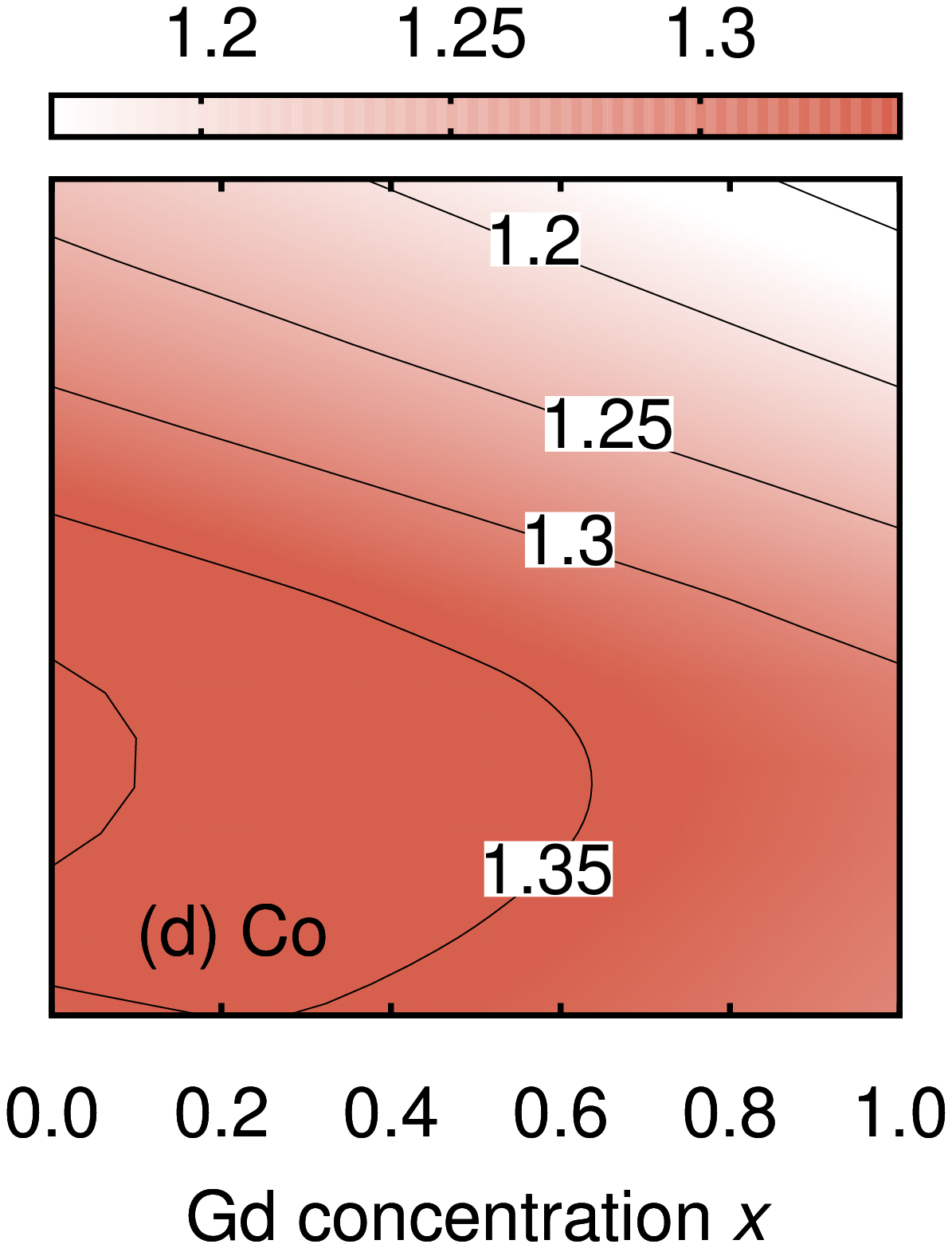}
\caption{\label{fig:mom} 
Magnetic moments in the ground state ($T$~=~0~K) for the \ygdfeco{} system calculated using the SPR-KKR code.
}
\end{figure*}

%
In Fig.~\ref{fig:Tc} we presents a Curie temperature dependence of Gd and Co concentrations for \ygdfeco{}
determined from Monte Carlo simulations with parameters from first-principles calculations.
For comparison, the four curves (red, green, blue, and yellow) are plotted based on experimental results~\cite{guzdek_electrical_2012,buschow_magnetic_1970,burkov_magnetotransport_2003,burzo_paramagnetic_1978,kilcoyne_evolution_2000}.
Along the Co concentrations ($y$), our MC simulations preserve the characteristic Slater-Pauling behavior for T$_C$($y$). 
For \ygdfe{}, we reproduce the linear behavior of the dependence of T$_C$ on Gd concentration, please compare with the bottom panel of Fig.~\ref{fig:cross}. 
Our T$_C$ estimates agree relatively well with experimental results in all boundary regions except for the \yco{} neighborhood. 
The reason for the observed discrepancy is a combination of two factors: 
(a) the paramagnetic nature of the \yco{} remaining on the verge of satisfying the Stoner criterion $(N(E_F)I \simeq 0.9)$~\cite{gratz_physical_2001,wasilewski_electronic_2019}, and
(b) the previously raised problem of GGA's failure to correctly describe the magnetic ground state of Co~\cite{wasilewski_curie_2018}.
Leaving aside the overestimation of the results near \yco{}, the map obtained for a system containing three magnetic elements shows that a realistic first-principles $T_C$ analysis for complex magnetic materials is possible allowing, for example, to determine the concentrations of the individual components leading to a specific $T_C$ value, e.g. 300~K, as sought for magnetocaloric materials for use in magnetic coolers.

%
$T_C$ values were estimated from magnetic moments and exchange integrals calculated from first principles.
Figure~\ref{fig:JXC} shows the calculated exchange integrals as a function of normalized interatomic distance for considered ferrimagnetic boundary compositions. 
It is easy to see that in each case the dominant contribution to the Heisenberg Hamiltonian comes from first-neighbor interactions at the Fe/Co sites. 
We verified that restricting the MC simulation to this dominant exchange interaction would lead to an underestimation of the Curie temperature on the order of hundreds of kelvins compared to simulations that include other exchange integrals.
The high values of first-neighbor Fe-Fe interactions observed in \yfe{}, further increase in \gdfe{}, as reflected by the linear increase in $T_C$ observed with increasing Gd concentration in \ygdfe{} alloys, compare with Fig.~\ref{fig:Tc} and the bottom panel of Fig.~\ref{fig:cross}. 
Similar conclusions can be drawn from the comparison of Fe-Fe and Co-Co first-neighbor interactions for Gd compounds.

%
In addition to the values of exchange interactions just discussed, the second factor that significantly affects the $T_C$ values obtained are the magnetic moments.
Figure~\ref{fig:mom} presents the calculated magnetic moments in the ground state ($T$~=~0~K) for the \ygdfeco{} system.
While the atom-resolved values of magnetic moments do not differ by more than $\sim 0.2~\mu_B$ with concentrations, the total magnetic moment varies from about $-4~\mu_B$ to $+4~\mu_B$.
This range is due to the antiparallel alignment of moments at the Fe/Co and Gd sites.
The calculated values of the magnetic moments are about 2.2 and 1.3~$\mu_B$ for Fe and Co and about -6.6~$\mu_B$ for Gd.
These results are in good agreement with our previous theoretical results for the systems \ygdco{}~\cite{pierunek_normal_2017} and \yfeco{}~\cite{wasilewski_electronic_2019}.
The calculated total magnetic moments are also in good agreement with the experimental results for the systems \ygdco{}~\cite{pierunek_normal_2017} and \yfeco{}~\cite{piercy_evidence_1968}.
Interestingly, for certain concentrations we observe a zero total moment, implying a complete compensation of the opposite moments present on the different elements.
We would like to recall that the results in the upper left corner of the magnetic moment map, describing the nearest neighborhood of \yco{}, are incorrect because \yco{} is in fact a magnetically disordered phase whose ground state is not correctly described in GGA.

%
For the positive (red) region of the total magnetic moment, due to the presence of uncompensated opposite magnetic moments, the relation $M$($T$) shows a positive slope at low temperatures, see Fig.~\ref{fig:JXC}(d).
For the example  Y$_{0.8}$Gd$_{0.2}$Co$_2$ concentration, our MC predicted increase in magnetization at low temperatures is confirmed by experimental (zero-field cooled) results~\cite{pierunek_normal_2017}.
For considered compounds with large positive total magnetic moment, 
we expect a large inverse magnetocaloric effect to occur in a region with positive $M$($T$) slope. 
The physics responsible for the shape of $M$($T$) curves can be understood by considering exchange integrals for different pairs of atoms. 
In Fig.~\ref{fig:JXC} we see that the 4\textit{f}-4\textit{f} and 4\textit{f}-3\textit{d} interactions are much weaker than the 3\textit{d}-3\textit{d} interactions therefore the thermal disorder first affects the 4\textit{f} sublattice, while the 3\textit{d} sublattice remains (at least relatively) ordered, which means that the magnetic moments of the 4\textit{f} sublattice no longer compensate, or compensate to a much lesser extent, the 3\textit{d} moments, hence the initial increase in magnetization. 
For temperatures higher than the temperature for which the magnetization maximum occurs in Fig.~\ref{fig:JXC}(d), the simulated material begins to behave as a ferromagnet.

\begin{figure*}[t]
\centering
\includegraphics[clip,height=0.57\columnwidth]{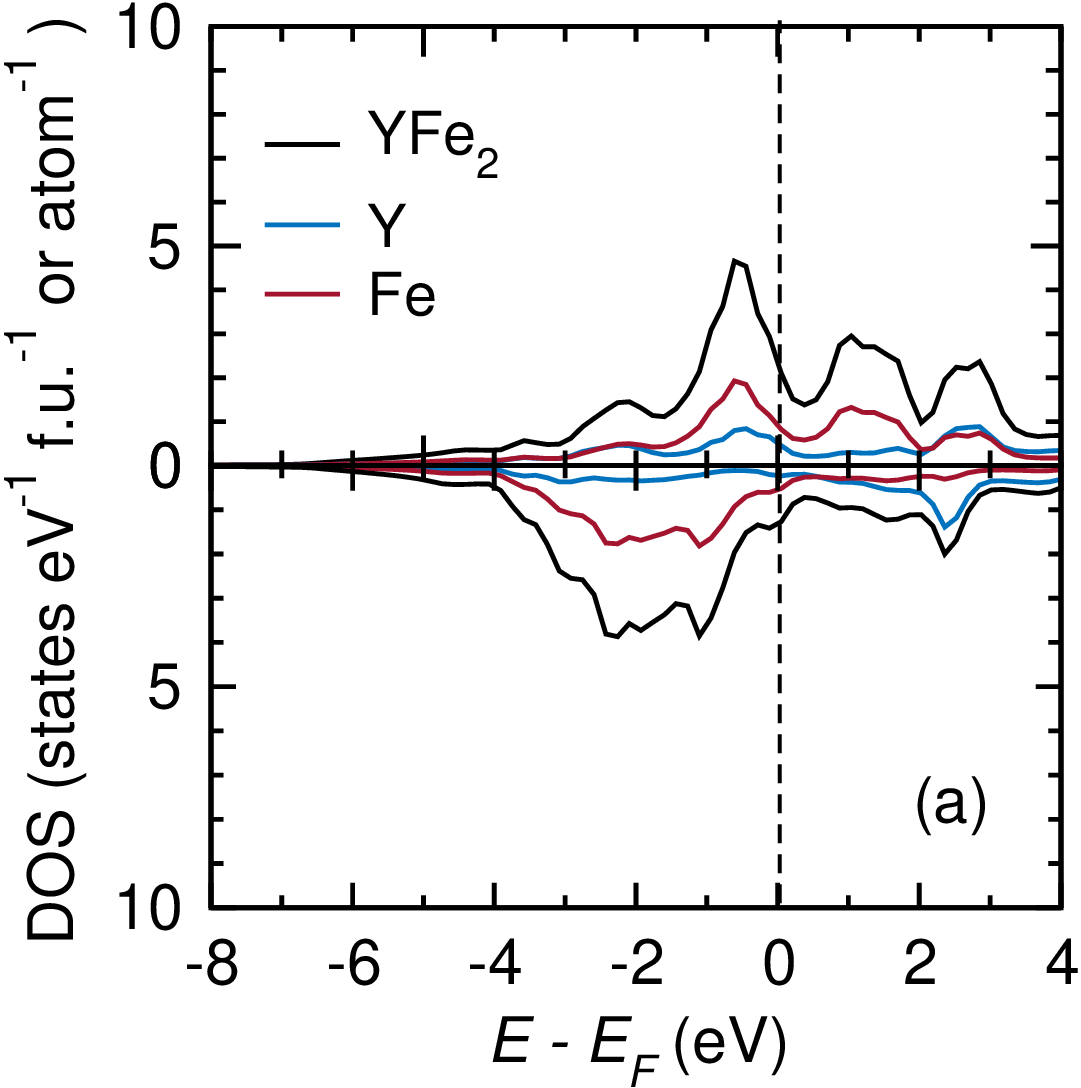}\hspace{1mm}
\includegraphics[clip,height=0.56\columnwidth]{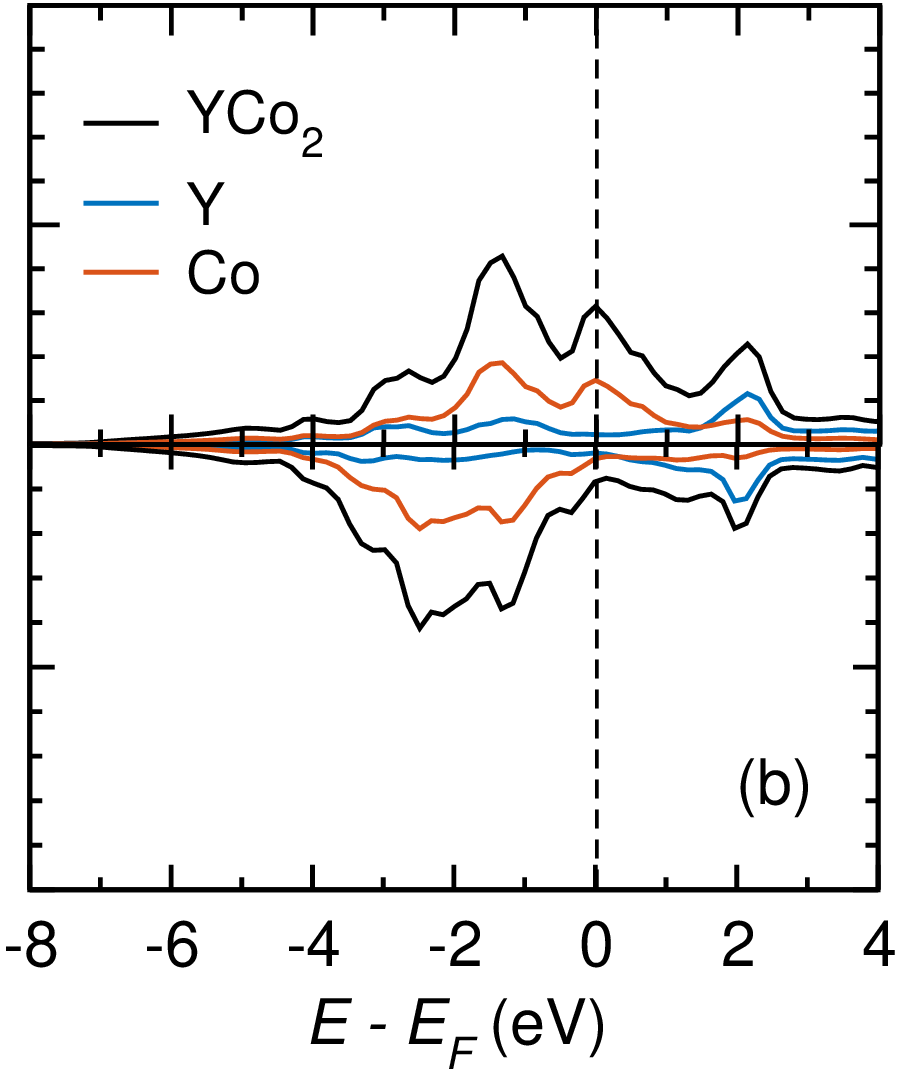}\hspace{1mm}
\includegraphics[clip,height=0.57\columnwidth]{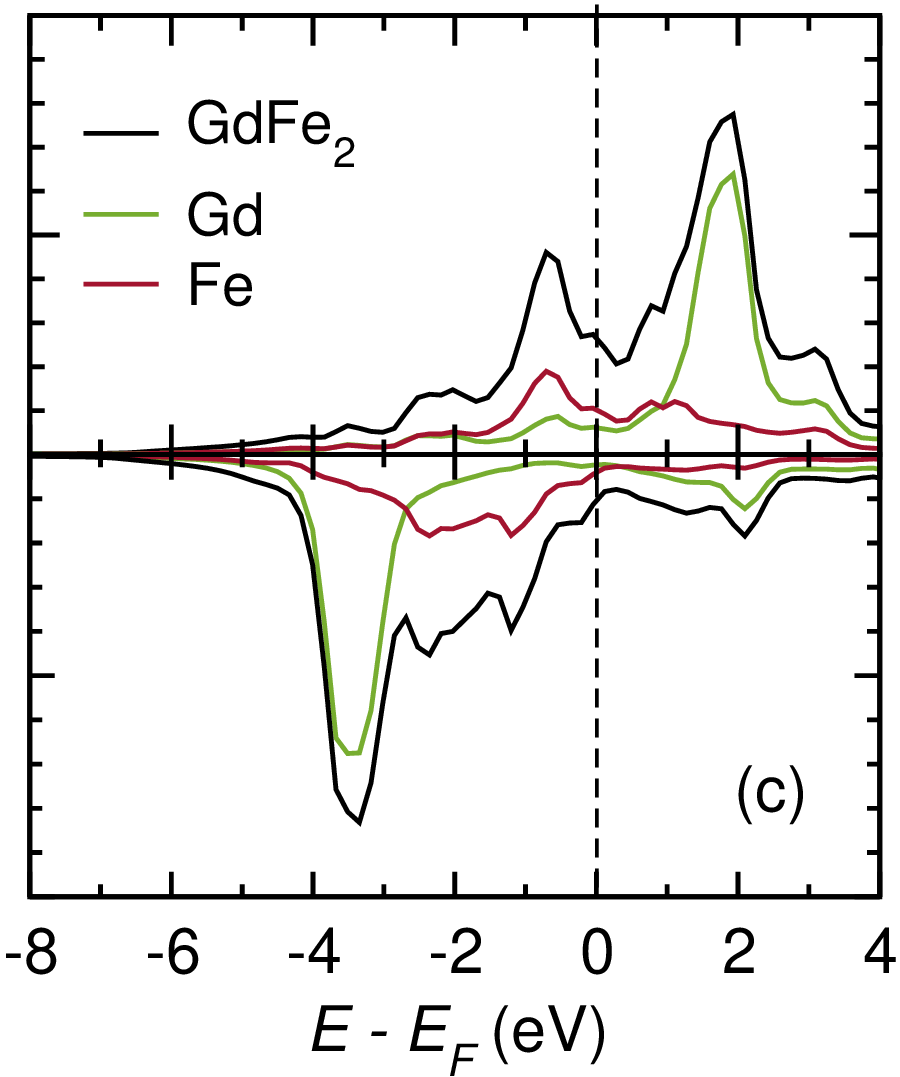}\hspace{1mm}
\includegraphics[clip,height=0.57\columnwidth]{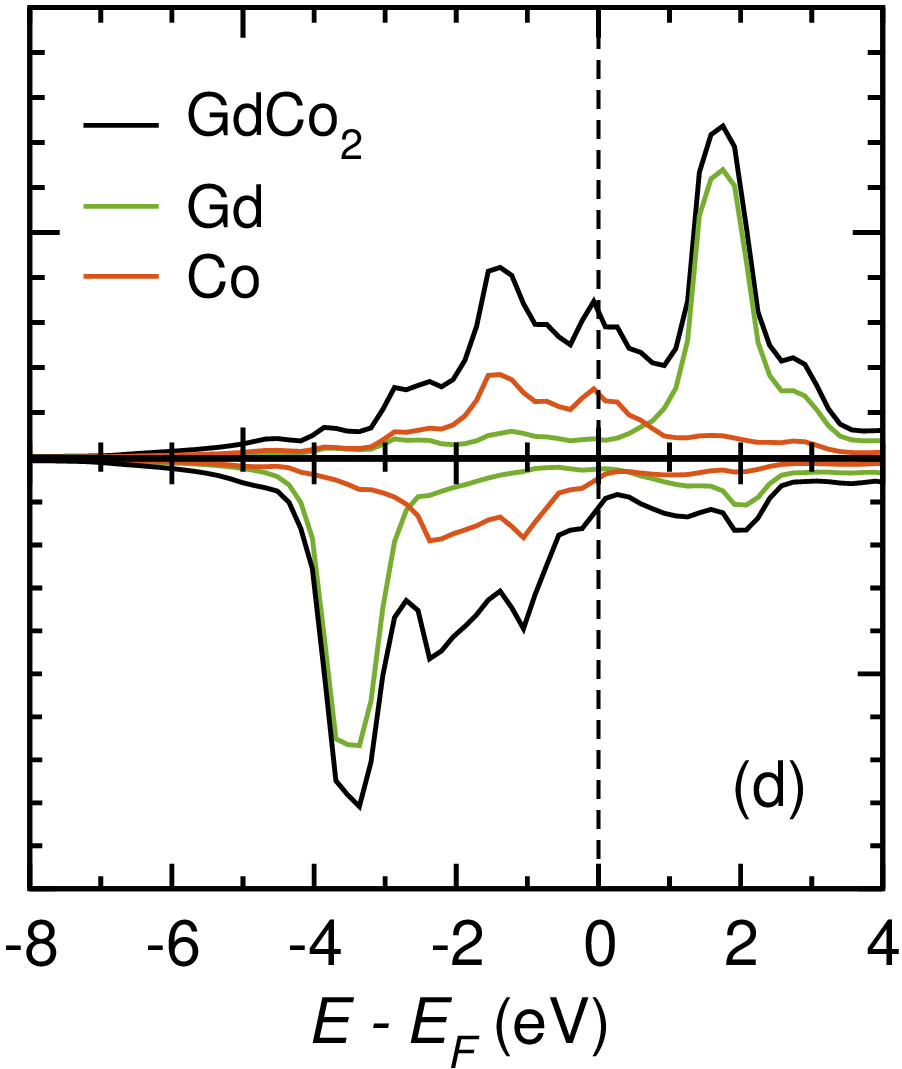}
\caption{\label{DOS} 
Densities of states (DOS) of \yfe{}, \yco{}, \gdfe{}, and \gdco{} as calculated with the SPR-KKR code.}
\end{figure*}

%
Figure~\ref{DOS} shows the valence band DOS for four stoichiometric compositions of the \ygdfeco{} system. 
All panels of Fig.~\ref{DOS} show results with spin polarization that is proportional to the values of calculated magnetic moments.
As we said, the magnetic moments are about 2.2, 1.3, and -6.6~$\mu_B$ for Fe, Co, and Gd.
In each case, the dominant contribution to the valence band comes from the 3$d$ orbitals of the 3$d$ elements (Fe and Co).
For compounds with Gd, we observe an almost completely occupied one spin channel of the Gd~4$f$ orbital and an almost empty second spin channel.
The 4$f$ bands are also much more localized than the 3$d$ bands.
The location of the occupied 4$f$ bands is about -3.5~eV below the Fermi level.
The difference in DOS between systems containing Co and Fe is mainly due to  filling of the valence band with an extra electron for each Co atom relative to the Fe atoms.
Similarly, as we mentioned earlier when discussing the results of Curie temperature calculations, the obtained ferromagnetic solution for \yco{} should be treated with caution in view of the experimentally found paramagnetic ground state for this phase.

\section{Summary and Conclusions}
In this paper, we theoretically described the Curie temperature dependence of the \ygdfeco{} pseudo-binary system using the combined Monte Carlo and first-principles methods. 
The calculation results agree well with the previous measurements except for the vicinity of the \yco{} system. 
For the latter, instead of paramagnetic ground state, we obtained a ferromagnetic one. 
For the \ygdfe{} subsystem we reproduced the linear dependence of T$_C$ on Gd concentration, and for the \yfeco{} and \gdfeco{} subsystems we reproduced the characteristic Slater-Pauling-like dependence of T$_C$ on concentration. 
The results presented here confirm the ability to efficiently predict Curie temperatures for magnetic systems containing up to three different magnetic elements simultaneously.
Furthermore, the first-principles results show that the largest contribution to the Heisenberg Hamiltonian comes from Fe/Co nearest-neighbor exchange interactions.
For the \ygdfeco{} system, we have also shown how the occurrence of a compensation point with an effective zero magnetic moment depends on the concentration of Co and Gd.
In addition, using the example of the Y$_{0.8}$Gd$_{0.2}$Co$_2$ ferrimagnetic phase, we have shown that the model used allows to predict non-trivial magnetization-temperature dependence.

\section*{Acknowledgements}
MW acknowledges the financial support of the National Science Centre Poland under the decision DEC-2018/30/E/ST3/00267.
We acknowledge the financial support from the Foundation of Polish Science grant HOMING.
The HOMING programme is co-financed by the European Union under the European Regional Development Fund.
The computations were performed on resources provided by the Poznan Supercomputing and Networking Center (PSNC).
We thank Paweł Leśniak and Daniel Depcik for compiling the scientific software and administration of the computing cluster at the Institute of Molecular Physics, Polish Academy of Sciences.
We also thank Zbigniew \'S{}niadecki for his comments and fruitful discussion.

\bibliography{ygdfeco2}    

\end{sloppypar}
\end{document}